\documentclass[12pt]{article}

\usepackage[T1]{fontenc}
\usepackage[utf8]{inputenc}
\usepackage{authblk}
\usepackage{graphicx}
\usepackage{mathtools}
\usepackage{url}
\usepackage{lipsum}
\usepackage{xcolor}

\let\origthanks\thanks
\renewcommand\thanks[1]{\begingroup\let\rlap\relax\origthanks{#1}\endgroup}

\title{Efficient Bayesian estimation of a non-Markovian Langevin model driven by correlated noise}

\author{Clemens Willers\thanks{clemens.willers@uni-muenster.de, ORCID ID: 0000-0001-5777-0514}}
\author{Oliver Kamps\thanks{okamp@uni-muenster.de, ORCID ID: 0000-0003-0986-0878}}
\affil{Center for Nonlinear Science (CeNoS), Westf{\"a}lische Wilhelms-Universit\"at M\"unster, Corrensstr.\ 2, 48149 M\"unster, Germany}
\date{\today}

\begin{document}

\maketitle

\begin{abstract}
Data-driven modeling of non-Markovian dynamics is a recent topic of research with applications in many fields such as climate research, molecular dynamics, biophysics, or wind power modeling. In the frequently used standard Langevin equation, memory effects can be implemented through an additional hidden component which functions as correlated noise, thus resulting in a non-Markovian model. It can be seen as part of the model class of partially observed diffusions which are usually adapted to observed data via Bayesian estimation, whereby the difficulty of the unknown noise values is solved through a Gibbs sampler. However, when regarding large data sets with a length of $10^6$ or $10^7$ data points, sampling the distribution of the same amount of latent variables is unfeasible. For the model discussed in this work, we solve this issue through a direct derivation of the posterior distribution of the Euler-Maruyama approximation of the model via analytical marginalization of the latent variables. Yet, in the case of a nonlinear noise process, the inverse problem of model estimation proves to be ill-posed and still numerically expensive. We handle these complications by restricting the noise to an Ornstein-Uhlenbeck process, which considerably reduces the ambiguity of the estimation. Further, in this case, the estimation can be performed very efficiently if the drift and diffusion functions of the observed component are approximated in a piecewise constant manner. We illustrate the resulting procedure of efficient Bayesian estimation of the considered non-Markovian Langevin model by an example from turbulence.
\end{abstract}

\vspace{2pc}
\noindent{\it Keywords}: Langevin model, correlated noise, non-Markov dynamics, Bayesian estimation, time series analysis, turbulence

\section{Introduction\label{sec:intro}}

In recent years, the data driven analysis of numerical systems receives growing interest in many fields of science. The increasing availability of data encourages the development of new methods and approaches for various applications in finance, ecology, climate research, or neuroscience, to name just a few. Specifically, the modeling of time series that describe stochastic dynamics is an important task. Here, Langevin models are a powerful model class covering many stochastic characteristics \cite{ReviewFriedrichPeinke}.

The basic version of these models is the standard Langevin equation (SLE), in which the dynamics is defined by a drift and a diffusion function and driven by Gaussian white noise. As a consequence of the uncorrelated noise, the SLE represents a Markov process, that is, the modeled dynamics exhibits no memory. The Markov property is a rough simplification for a general stochastic process. It is frequently assumed as an approximation. However, there are also many real-world systems in which Markov modeling is not sufficient, especially in the context of relatively high sampling rates, i.e., if the system's memory time is larger than the time scale of interest.

Stochastic models incorporating memory effects are a recent topic of research in many different fields such as (paleo-)climate \cite{GreenlandIce,ElNino_DelayDE,Memory_Climate} and power grid \cite{Pesch, Katrin2DLangevin} modeling. Another important example is the projection of a full multi-dimensional molecular dynamics (MD) simulation of, e.g., biomolecular folding onto low-dimensional reduced coordinates (also termed coarse-graining). Hereby, memory effects come into account, if essential and non-essential coordinates cannot be strictly separated \cite{GLE_MD_1,GLE_MD_2,Review_GLE_MD}. Further, the anticipation of critical transitions is a current topic of research with applications in many different fields. Here, the considered early-warning indicators rely on a description of the data by a non-stationary, linear SLE. Yet, a recent study shows that this approach is likely to cause false alarms if the assumption of uncorrelated noise is not fulfilled \cite{Boers}, which is probable in real world examples such as climate data \cite{Scheffer}. A proposed new indicator is based on a non-Markovian Langevin model incorporating correlated noise \cite{Boers}.

To obtain a Langevin model representing non-Markovian processes, two approaches are possible. Firstly, the SLE can be augmented with a memory integral yielding a stochastic delay differential equation, the generalized Langevin equation (GLE) \cite{Review_GLE_MD,GLE_PRE,GLE_climate,GLE_PRX,GLE_financial}. The GLE dynamics is driven by correlated noise if the fluctuation-dissipation theorem is valid \cite{Review_GLE_MD}. Otherwise (e.g. far from thermodynamical equilibrium) or as an approximation, a GLE model with white noise can be employed \cite{GLE_PRE,GLE_nonEquilibrium,GreenlandIce}.

Alternatively, the SLE can be extended by additional components leading to a higher-dimensional SLE driven by white noise. Hereby, the additional coordinates are \textit{hidden}, i.e., regarding a physical system, their values cannot be measured (these are also termed \textit{latent variables}). This type of model is sometimes called \textit{partially observed diffusion}. As a result, while the dynamics is Markovian with respect to all coordinates, it is non-Markovian with respect to the measurable coordinates only. The hidden components act as a memory of the observed process. Both approaches are connected in the sense that the GLE can be approximated by an extended SLE, which is used to avoid the expensive computation of both the memory integral and the correlated noise during integration of the model \cite{nonMarkov_by_highDimensionalMarkov_1,nonMarkov_by_highDimensionalMarkov_2,nonMarkov_by_highDimensionalMarkov_3,GLE_Lei1,GLE_Lei2,closureModels,GLE_BayesianOPTIMIZATION,ComputingGLEsimulation}.

In this work, we follow the latter approach through defining a 2-dimensional Langevin model whose hidden second component functions as a correlated noise process. This is a straightforward extension of the 1-dimensional SLE which involves memory effects. It has also been discussed in, e.g., references \cite{Katrin2DLangevin,BerndOUnoise,OUdrivenbyOU,directEst,Boers,ColoredNoiseModelBsp1,ColoredNoiseModelBsp2,ColoredNoiseModelBsp3,ColoredNoiseModelBsp4}, which include applications in various fields. In many of these works, the noise component is an \textit{Ornstein-Uhlenbeck (OU) process} which means that the drift function is linear and the diffusion function is constant (\textit{additive noise}). In references \cite{Boers,OUdrivenbyOU}, both components of the model are OU processes. A more complicated model with a nonlinear drift function in the noise process is used in, e.g., reference \cite{ColoredNoiseModelBsp2}. We start our discussion with the most general case, i.e., with arbitrary drift and diffusion functions. The considered model (for which we use the abbreviation HLE: \textbf{h}idden noise \textbf{L}angevin \textbf{e}quation) reads as follows:

\begin{subequations}
\label{HLEgeneral}
\begin{align}
\dot{X}_t &= D^{(1)}_x(X_t) + \sqrt{D^{(2)}_x(X_t)}\, Y_t\\
\dot{Y}_t &= D^{(1)}_y(Y_t) + \sqrt{D^{(2)}_y(Y_t)}\,\eta_t.
\end{align}
\end{subequations}

\noindent Hereby, $\eta_t$ is Gaussian white noise ($\langle\eta_t,\eta_{t'}\rangle = \delta(t,t')$), that is, $\eta_t$ is uncorrelated, whereby $Y_t$ is correlated. As mentioned, $X_t$ represents the observed process and the values of $Y_t$ are latent variables. There are two views on this system. On the one hand, it could be regarded as a 2-dimensional SLE in which the first component exhibits no diffusion function since it is not affected by the white noise process $\eta_t$. On the other hand, the process $Y_t$ is interpreted as the noise driving the observed process $X_t$. From this perspective, $D^{(2)}_x$ is the corresponding diffusion function. To emphasize the interpretation of the model, we call $D^{(1)}_x$ and $D^{(1)}_y$ \textit{drift functions} and $D^{(2)}_x$ and $D^{(2)}_y$ \textit{diffusion functions}, as other authors do. Besides, we consider $X_t$ and $Y_t$ as 1-dimensional and the model as stationary, i.e. the drift and diffusion functions do not depend on time. To examine the stationarity of a measured process, a moving-window analysis can be applied. Accordingly, a non-stationary process can be modeled by assigning different models to different windows in time \cite{ReviewFriedrichPeinke} (as it is done in the analysis of critical transitions mentioned above).

The numerical integration of the HLE (which will also be important for Bayesian estimation) can be realized through the Euler-Maruyama scheme (we use It\^{o} calculus), which reads \cite{KloedenPlaten}

\begin{subequations}
\label{EulerMaruyama}
\begin{align}
X_{i+1} &= X_i + D^{(1)}_x(X_i)\,\Delta t + \sqrt{D^{(2)}_x(X_i)}\, Y_i\,\Delta t\\
Y_{i+1} &= Y_i + D^{(1)}_y(Y_i)\,\Delta t + \sqrt{D^{(2)}_y(Y_i)}\,\sqrt{\Delta t}\, N_i
\end{align}
\end{subequations}

\noindent for a time step $\Delta t$ and a sequence of independent Gaussian random numbers $N_i\sim\mathcal{N}(0,1$). Due to the correlation of the noise $Y_t$, we have the relation

\begin{equation}\label{MP_HLE}
p(x_{i+1} | x_i, x_{i-1}, ..., x_0) = p(x_{i+1} | x_i, x_{i-1}),
\end{equation}

\noindent for the conditional probability of a value $x_{i+1}$ at time $i+1$ of the time series, given all previous values, as we derived in a previous work \cite{directEst}. This indicates the incorporated memory, which does not exist in the SLE model, in which case we would find the term $p(x_{i+1} | x_i)$ on the right-hand side of equation (\ref{MP_HLE}). Here as well as in the following, we regard all processes as discrete in time, because we deal with sampled measurements or numerical integrations. We use capitals for random variables and lowercases for their realizations.

Usually, both the SLE and the HLE are used to describe the dynamics of a macroscopic observable. Hereby, in principal, their specific form can be deduced from high-dimensional microscopic equations of motion by a projection operator according to the Mori-Zwanzig formalism \cite{GrabertsBuch}. Yet, in many examples, this direct derivation is not accessible, because the projection or the microscopic theory itself is not achievable in an explicit form. In these cases, one aims to estimate a model on the basis of simulated or measured data \cite{ReviewFriedrichPeinke}. Among various estimation techniques, Bayesian estimation has the advantage of providing a simple and flexible framework which allows the estimation of both optimal model parameters and their credible intervals. It is based on the \textit{posterior} which is the probability distribution of the parameters given the observed data. An estimate can be obtained by maximizing the posterior (MAP, \textit{maximum a posteriori} estimate); credible intervals of individual parameters can be defined through the corresponding marginal posterior distributions (for a detailed explanation, see section \ref{sec:Bayes}).

In this work, we discuss Bayesian parameter estimation of the HLE as defined in equation (\ref{HLEgeneral}) given an observed time series of $10^6$ or more data points. Here, the hidden values are a special difficulty. The literature provides different methods for the class of partially observed diffusion models \cite{POD_ReviewSorensen}. The main difficulty occuring there results from a sparse time-discrete observation of the process due to which the transition probability between consecutive values is not available in closed form (therefore, also the term \textit{discretely observed diffusion} is used). It is solved by \textit{data-augmentation}, i.e., by the introduction of \textit{diffusion bridges} as latent variables between every two consecutive observed values \cite{POD_ReviewSorensen,MCMC3,POD_vdMeulen,DOD_vdMeulen}. In our applications, we assume that we do not face such discretization errors, i.e., that the sampling of the observed process is fine enough for the Euler-Maruyama approximation of the estimated model to be valid. However, a few authors also discuss the situation in which only part of the components of the diffusion model are observed as in the HLE model (in this case, also the term \textit{incompletely observed diffusion} is used). Most of them follow a Bayesian Markov Chain Monte Carlo (MCMC) approach to the estimation of the model parameters \cite{JensenPHD,POD_vdMeulen,MCMC3,POD_WuNoe,POD_Beskos}. Hereby, the posterior is investigated through a \textit{Gibbs sampler}, which iteratively samples both model parameters and latent variables in different steps. Some authors assume an invertible \cite{JensenPHD,POD_vdMeulen} or constant \cite{POD_WuNoe} diffusion matrix, which does not apply in our model (if regarded as a 2-dimensional SLE). However, all methods share the difficulty that Gibbs sampling of latent variables is numerically expensive. In reference \cite{MCMC3}, for instance, the lengths of the regarded data sets are in the order of magnitude of $10^2$. Sampling the distribution of $10^6$ or more hidden variables, as it would be necessary in our applications, is impossible.

Generally, the sampling of latent variables can be avoided in different ways. First, an MAP estimate can be obtained by the iterative \textit{expectation-maximization} (EM) algorithm \cite{EM0,POD_Beskos}, which asymptotically approaches a (local) maximum of the posterior. Yet, in the case of the HLE model, which does not exhibit a white noise term in the first component, the iterative steps of the EM algorithm would only change the parameters of the second component. The parameters of the first component would keep the values of the initial guess.
Second, \textit{approximate Bayesian computation} (ABC) \cite{ABC0,ABC1} can be applied to approximate the posterior through simulations of the model which are compared to the observed data via characterizing summary statistics. This facilitates both sampling (though we expect that this is computationally expensive in our case) and optimization (see also references \cite{ABC_MAP1,ABC_MAP2}) of the posterior. However, in practice, it is unfeasible to define summary statistics which are \textit{sufficient}, i.e., which contain as much information about the underlying model parameters as the simulation itself (a good candidate is the transition probability density of equation (\ref{MP_HLE}), but this too can only be estimated). Therefore, part of the information of the data is lost and the results are biased to a certain extend.

In the case of the HLE model, it is possible to solve the difficulty of the latent variables in a much more direct way, namely by marginalization which we perform in section \ref{sec:Propagator}. Thereby, we obtain a posterior distribution which only depends on the model parameters and the observed data and Bayesian estimation can be performed straightforwardly.

During application of Bayesian estimation in the case of the general HLE model incorporating a nonlinear noise process, we observe two difficulties. First, as we will show in section \ref{sec:HLE-OU}, the estimation forms an ill-posed inverse problem, i.e., its solution is not unique. Second, the evaluation of the posterior still is numerically expensive when dealing with large data sets. As a consequence, while MAP estimation is feasible, MCMC sampling of the posterior proves problematic. We do not want to address these issues concerning the general model in this work. Instead, in a subsequent step, we discuss Bayesian estimation of the special model incorporating OU noise, in which we circumvent these difficulties:

\begin{subequations}
\label{HLE}
\begin{align}
\dot{X}_t &= D^{(1)}_x(X_t) + \sqrt{D^{(2)}_x(X_t)}\, Y_t\\
\dot{Y}_t &= -\frac{1}{\theta}\,Y_t + \sqrt{\frac{1}{\theta}}\,\eta_t.
\end{align}
\end{subequations}

\noindent We use the abbreviation HLE-OU for this special model. It involves memory effects through the non-zero correlation time of the noise, which is defined by the parameter $\theta$. These effects can, for example, be observed in the autocorrelation function (ACF) of the process $X_t$. Here, the different time scales of the different components of the model can lead to a curve of the ACF which is concave for small time increments \cite{directEst}. This phenomenon is often observed in examples where memory effects are essential such as in biomolecular folding \cite{GLE_MD_2} or wind power modeling \cite{Pesch,LevystableSuperstatistics}. Generally, an ACF that does not decay single exponentially in time (but, e.g., multi exponentially or according to a power law) is an indication for memory effects or colored noise \cite{GLE_nonEquilibrium}. The HLE-OU model can be applied in various fields, as shown for instance in references \cite{Katrin2DLangevin,OUdrivenbyOU,Boers,ColoredNoiseModelBsp1,ColoredNoiseModelBsp3,ColoredNoiseModelBsp4}.

The restriction of the noise to an OU process considerably reduces the ambiguity of the estimation problem (for a detailed explanation, see section \ref{sec:HLE-OU}). Furthermore, the estimation of the HLE-OU can be performed very efficiently through an idea which was first employed by Kleinhans in the case of the SLE \cite{Kleinhans}: If the functions $D^{(1)}_x$ and $D^{(2)}_x$ are approximated piecewise constantly, the crucial characteristics of a measured time series can be represented by a few coefficients derived from conditional sums of specific increments. Taking only these coefficients into account, the posterior can be evaluated much more efficiently than by considering every single value of the data set in an unsorted way. For time series of length $10^6$ or more, this makes a huge difference in the computation time of Bayesian estimation. Actually, the computation time of optimization and sampling of the posterior gets independent of the length of the data set. This efficient realization of the estimation is only possible in the special case of OU noise.

For a Langevin process driven by OU noise, the literature provides a few estimation techniques. First, there is a method developed by Lehle et al. \cite{BerndOUnoise}. It only works in the case that the time scale of the noise process $Y_t$ is small compared to the time scale of the measured process $X_t$, because it relies on a perturbative approach. A second method can be found in a previous work of ours \cite{directEst}. It does not include the limitation concerning the time scales. Yet, it is restricted to examples in which the measured dynamics can be approximated by a linear model to a certain extend. Further, Bercu et al. \cite{OUdrivenbyOU} discuss maximum likelihood estimation of an OU process which is driven by OU noise. The model only includes two free parameters.
All these methods are applicable to large data sets. However, unlike Bayesian estimation, they do not allow the determination of credible intervals. Further, as mentioned, they are all subject to specific restrictions, which do not apply to the method proposed in this work.

In section \ref{sec:exampleTurb}, we present an application of Bayesian estimation of the HLE-OU model to an example from turbulence, in which this non-Markovian model clearly outperforms the Markovian SLE.

This work is organized as follows. First, we provide a brief introduction to Bayesian estimation (section \ref{sec:Bayes}). Second, we derive the posterior of the HLE model, which Bayesian estimation is based on (section \ref{sec:Propagator}), via marginalization. Next, we examine the resulting method through a synthetic test example (section \ref{sec:exampleSynth}). Hereby, we observe that Bayesian estimation of the HLE model forms a numerically expensive ill-posed inverse problem. Therefore, in the following, we regard the special HLE-OU model, in which the ambiguity of the estimation is significantly reduced (section \ref{sec:HLE-OU}). Further, based on the ideas of direct estimation of the SLE (which we explain in section \ref{sec:directEstSLE}), we implement Bayesian estimation of the HLE-OU in an efficient way (section \ref{sec:Efficient}). With a data set from turbulence, we present an application of this approach to non-Markovian modeling (section \ref{sec:exampleTurb}). Finally, we discuss our findings (section \ref{sec:summary}).

\section{Bayesian estimation}\label{sec:Bayes}

The Bayesian approach \cite{Toussaint,BayesianDataAnalysis} provides a very simple and flexible framework for the parameter estimation of a model. A further advantage is that it allows to determine credible intervals for the estimated parameters in a straightforward way. The idea is to regard the model parameters as random variables and to base the estimation on the probability of a set of parameters $\vartheta=(\vartheta_1, ..., \vartheta_{N_{\text{par}}})$ given the observed data $\xi=(x_0, x_1, ..., x_{N_{\text{data}}})$. This probability distribution is called the \textit{posterior} $p(\vartheta|\xi)$.

\noindent There are various possibilities of defining best parameter values based on the posterior. We search for its maximum to find estimates for the parameters $\hat{\vartheta}$ (maximum a posteriori estimation, MAP): $\hat{\vartheta} := \text{argmax}_{\vartheta}\, p(\vartheta|\xi)$. Instead, the mean of the posterior could be considered or the whole shape of the posterior could be investigated via an MCMC analysis. The latter approach is advantageous, if the posterior is, e.g., a bimodal or asymmetric distribution. Credible intervals for estimated parameters can be evaluated through the marginal distributions of the posterior. For a parameter $\vartheta_l$, the corresponding marginal distribution of the posterior is defined as (the integration is facilitated by the MCMC algorithm)

\begin{align}
p(\vartheta_l|\xi) = \int\hdots\int p(\vartheta|\xi)\,d\vartheta_1 \hdots d\vartheta_{l-1}d\vartheta_{l+1}\hdots d\vartheta_{N_{\text{par}}}.
\end{align}

Bayes' theorem allows the derivation of the posterior yielding a product of likelihood $p(\xi|\vartheta)$ and prior $p(\vartheta)$:

\begin{align}\label{BayesTheorem}
p(\vartheta | \xi) \propto p(\xi | \vartheta) \, p(\vartheta).
\end{align}

\noindent For a given model, the prior involves general information about (physically) reasonable parameter values. The likelihood can be calculated (possibly approximately or numerically) via the definition of the model. We show in section \ref{sec:Propagator} how this can be achieved for the HLE. Hereby, we tackle the difficulty of the unknown values of the second component of the model by marginalization.

\section{Derivation of the posterior}\label{sec:Propagator}

In the HLE model (cf. equation (\ref{HLE})), the model parameters $\vartheta=(\vartheta_1, ..., \vartheta_{N_{\text{par}}})$ are chosen according to the parameterization of the functions $D^{(1)}_x$, $D^{(2)}_x$, $D^{(1)}_y$, and $D^{(2)}_y$. This parameterization can be made by, e.g., polynomials, splines, or as piecewise constant functions (cf. section \ref{sec:Efficient}).

Let $\xi=(x_0, x_1, ..., x_{N_{\text{data}}})$ be an observed time series, which is sampled with a time step $\Delta t$ (i.e., $x_i$ is the value of $X_{i\Delta t}$ in the observed realization of the process). To carry out Bayesian estimation of the parameters, we have to derive the posterior $p(\vartheta|\xi)$. For technical reasons, both numerical optimization of the posterior and MCMC sampling require its logarithm (hereby, arithmetic under-/overflows are circumvented). Via Bayes' theorem (cf. equation (\ref{BayesTheorem})), we obtain

\begin{align}
\log p(\vartheta | \xi) \propto \log p(\xi | \vartheta) + \log p(\vartheta).
\end{align}

\noindent We define the prior $p(\vartheta)$ such that we involve the condition that the diffusion functions $D^{(2)}_x$ and $D^{(2)}_y$ should be positive. The likelihood $p(\xi|\vartheta)$ can be calculated through a seperation into individual steps (as usual in the context of time series \cite{GreenlandIce,NARMAX,SLE_Likelihood_BIC,Bayes_SLE,Kleinhans}):

\begin{align}
p(\xi|\vartheta) = p(x_0, ..., x_{N_{\text{data}}}|\vartheta) = \left( \prod_{i=0}^{{N_{\text{data}}}-1} p(x_{i+1} | x_i, x_{i-1}, ..., x_0,\vartheta) \right) p(x_0|\vartheta).
\end{align}

\noindent Now, we make use of the finiteness of the memory involved in the model, whereby we arrive at the short-term propagator $p(x_{i+1} | x_i, x_{i-1},\vartheta)$ (cf. equation (\ref{MP_HLE})). Hence, for the logarithm of the likelihood, we obtain (neglecting the term $\log p(x_1,x_0|\vartheta)$):

\begin{align}\label{loglikelihood}
\log p(\xi|\vartheta) = \sum_{i=1}^{{N_{\text{data}}}-1} \log p(x_{i+1} | x_i, x_{i-1},\vartheta).
\end{align}

In the following, we derive the short-term propagator $p(x_{i+1} | x_i, x_{i-1},\vartheta)$. Since the values of the noise process $Y_t$ are unknown, it has to be calculated through marginalization:

\begin{multline} \label{marginalization}
    p(x_{i+1}|x_i,x_{i-1},\vartheta) = \int p(x_{i+1},y_i|x_i,x_{i-1},\vartheta)\, dy_i \\
    = \int p(x_{i+1}|y_i,x_i,x_{i-1},\vartheta) p(y_i|x_i,x_{i-1},\vartheta)\, dy_i.
\end{multline}

\noindent We apply the Euler-Maruyama approximation of the HLE (cf. Eq.~(\ref{EulerMaruyama})) to derive the integrand. The condition in the second factor of the integrand leads to the value

\begin{align}
    y_{i-1} := \frac{x_i - x_{i-1} - D^{(1)}_x(x_{i-1})\Delta t}{\sqrt{D^{(2)}_x(x_{i-1})}\Delta t}.
\end{align}

\noindent Conditional on this value, the pdf of $y_i$ is a Gaussian with mean $\mu_y$ and variance $\sigma_y^2$, where

\begin{align}
    &\mu_y := y_{i-1} + D^{(1)}_y(y_{i-1})\Delta t &\text{and}&
    &\sigma^2_y := D^{(2)}_y(y_{i-1})\Delta t.
\end{align}

\noindent The first factor of the integrand corresponds to the Dirac delta function $\delta(g(y_i))$, whereby

\begin{align}
    g(y_i) := x_{i+1} - \left(x_i + D^{(1)}_x(x_i)\Delta t + \sqrt{D^{(2)}_x(x_i)}y_i\Delta t\right).
\end{align}

\noindent The zero of this function is

\begin{align}
    \hat{y}_i := \frac{x_{i+1} - x_i - D^{(1)}_x(x_i)\Delta t}{\sqrt{D^{(2)}_x(x_i)}\Delta t},
\end{align}

\noindent so that the integration with respect to $y_i$ yields

\begin{align}
    p(x_{i+1}|x_i,x_{i-1},\vartheta) = \frac{1}{|g'(\hat{y}_i)|}\frac{1}{\sqrt{2\pi\sigma_y^2}}\exp\left\{-\frac{(\hat{y}_i - \mu_y)^2}{2\sigma_y^2}\right\}.
\end{align}

\noindent Due to $|g'(y_i)| = \sqrt{D^{(2)}_x(x_i)}\Delta t$, we finally obtain the following expression for the short-term propagator:

\begin{multline}\label{propHNLE}
	p(x_{i+1}|x_i,x_{i-1},\vartheta) = \frac{1}{\sqrt{2\pi\sigma_{\vartheta}(x_i,x_{i-1})^2}} \\
	 \exp\left\{-\frac{\left(x_{i+1} - \mu_{\vartheta}(x_i,x_{i-1})\right)^2}{2\sigma_{\vartheta}(x_i,x_{i-1})^2}\right\},
\end{multline}

\noindent whereby

\begin{align}
    \mu_{\vartheta}(x_i,x_{i-1}) &:= x_i + D^{(1)}_x(x_i)\Delta t \nonumber\\
    &\qquad + \sqrt{D^{(2)}_x(x_i)}y_{i-1}\Delta t + D^{(1)}_y(y_{i-1})\sqrt{D^{(2)}_x(x_i)}\Delta t^2 \label{propHNLEa}\\
    \sigma_{\vartheta}(x_i,x_{i-1})^2 &:= D^{(2)}_y(y_{i-1}) D^{(2)}_x(x_i) \Delta t^3. \label{propHNLEb}
\end{align}

\noindent Hence, the logarithm of the likelihood function reads

\begin{multline}\label{likelihood_unsorted}
	\log p(\xi|\vartheta) = \sum_{i=1}^{N_{\text{data}}-1} \Biggl\{ -\frac12 \log \left(2\pi\sigma_{\vartheta}(x_i,x_{i-1})^2 \right) \\
	- \frac{\left(x_{i+1} - \mu_{\vartheta}(x_i,x_{i-1})\right)^2}{2\sigma_{\vartheta}(x_i,x_{i-1})^2} \Biggr\}.
\end{multline}

\noindent This completes the explicit expression for the posterior which facilitates Bayesian estimation of the HLE model.

\section{Synthetic example of HLE model estimation}\label{sec:exampleSynth}

To illustrate and examine the estimation of the general HLE model (\ref{HLEgeneral}), we consider the following synthetic example:

\begin{subequations}
\label{HLEsynthetic}
\begin{align}
\dot{X}_t &= X_t - X_t^3 + \sqrt{0.9 + 0.2X_t^2}\, Y_t\\
\dot{Y}_t &= -2Y_t^3 + \sqrt{0.2 + 0.6Y_t^2}\,\eta_t.
\end{align}
\end{subequations}

\noindent The drift function of the first component defines two stable fixed points and the correlated noise drives the dynamics back and forth. Hence, the distribution of the values of the process $X_t$ is bimodal. All drift and diffusion functions are nonlinear. We generate a time series of the process $X_t$ by Euler-Maruyama integration (\ref{EulerMaruyama}) with time step $\Delta t = 0.1$ and a length of $10^6$ data points. Based on this synthetic data set, we perform the estimation of an HLE model. Afterwards, we compare the result to the originally defined model. Hereby, we investigate several issues concerning the estimation of the general HLE model.

For the estimation, we parameterize the drift and diffusion functions by polynomials. In the first component of the model, we define

\begin{subequations}
\label{ParamSynth}
\begin{align}
D^{(1)}_x(x) &= \vartheta_0x + \vartheta_1x^2 + \vartheta_2x^3 \\
D^{(2)}_x(x) &= \vartheta_3 + \vartheta_4x + \vartheta_5x^2.
\end{align}
\end{subequations}

\noindent In the second component, we use the same polynomials. That is, in sum, we estimate 12 parameters. In the drift functions, we omit the constant term, which means that both $X_t$ and $Y_t$ have a fixed point at zero. For the noise process, this assumption is reasonable. For the observed component, it can be made without loss of generality in the symmetric case, i.e., if the fixed point of $X_t$ equals its mean $\langle X_t \rangle$. If $\langle X_t \rangle$ was nonzero, we would estimate the parameterized model for the process $\tilde{X}_t := X_t - \langle X_t \rangle$. In the case of an asymmetric distribution of the values of $X_t$, a constant term in the function $D^{(1)}_x$ could be necessary.

For reasons of efficiency, we perform the estimation in two steps. First, we estimate the linear model

\begin{subequations}
\label{HLEsimpleLinear}
\begin{align}
\dot{X}_t &= -\frac{1}{\vartheta^l_0}X_t + \sqrt{\vartheta^l_1}\, Y_t\\
\dot{Y}_t &= -\frac{1}{\vartheta^l_2}Y_t + \sqrt{\frac{1}{\vartheta^l_2}}\,\eta_t.
\end{align}
\end{subequations}

\noindent which reflects the most important basic properties of the data set: the time scales of the processes $X_t$ and $Y_t$ ($\vartheta^l_0$ and $\vartheta^l_2$) and the variance of $X_t$ (determined by $\vartheta^l_1$) (see \cite{directEst} for a detailed discussion of this model). Its estimation is well manageable, which means that the numerical optimization of the posterior is fast and that the result is (mostly) independent of the initial guess. Here, we start the Bayesian estimation with the neutral initial guess $(\vartheta^l_0, \vartheta^l_1, \vartheta^l_2) = (1,1,1)$. Next, we use the estimated linear model as an initial guess for the estimation of the general model with polynomial parameterization.

For the numerical optimizations, we apply a combination of the Powell \cite{Powell} and Nelder-Mead \cite{NelderMead} algorithms. According to our experience in the context of Bayesian estimation of Langevin models, the first algorithm is fast, while the second is precise. Therefore, using Powell in a first step and Nelder-Mead in a second step with the Powell result as initial guess results in an effective optimization.

\begin{figure}
  \includegraphics[width=0.49\hsize]{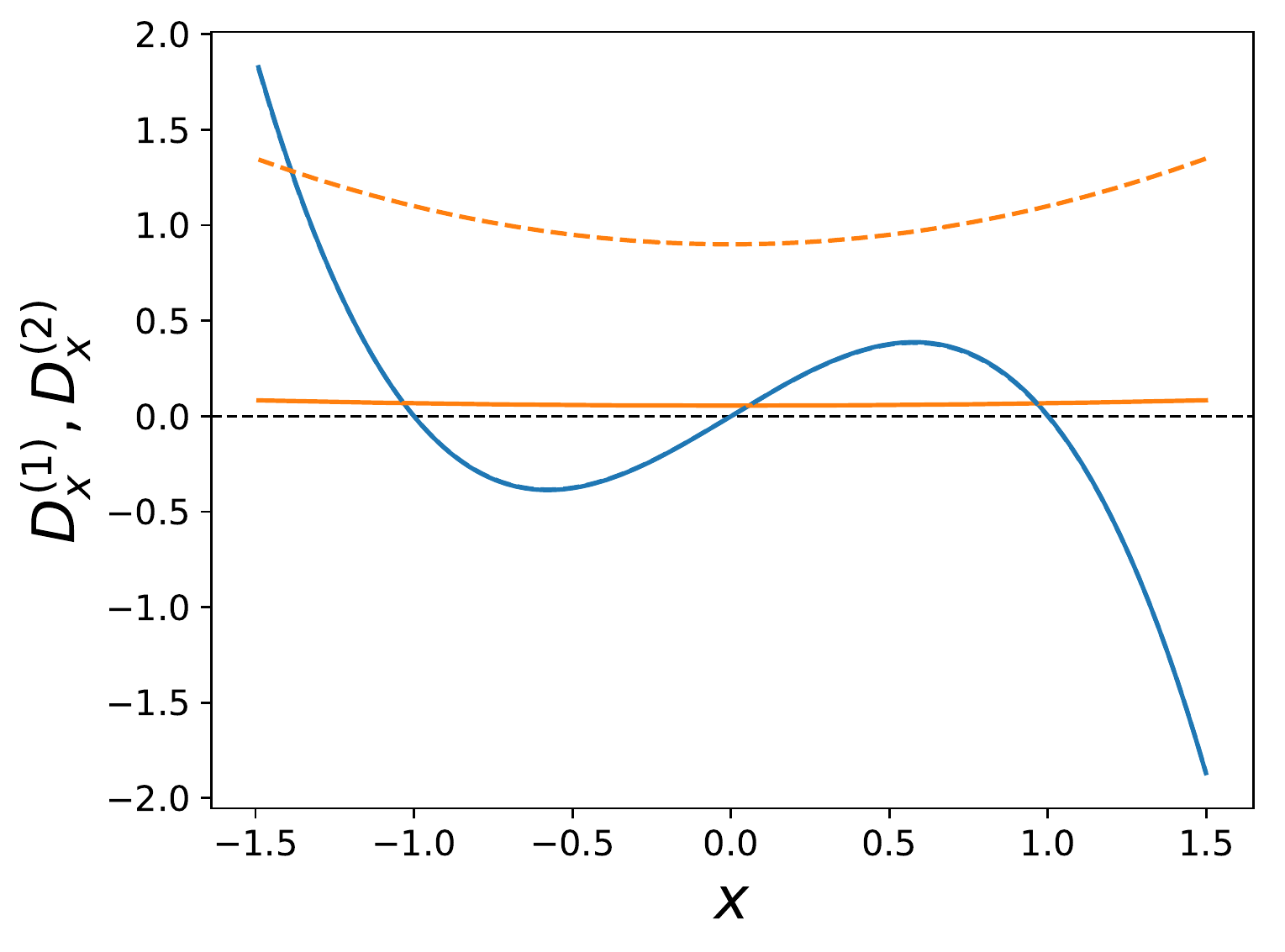}
  \includegraphics[width=0.49\hsize]{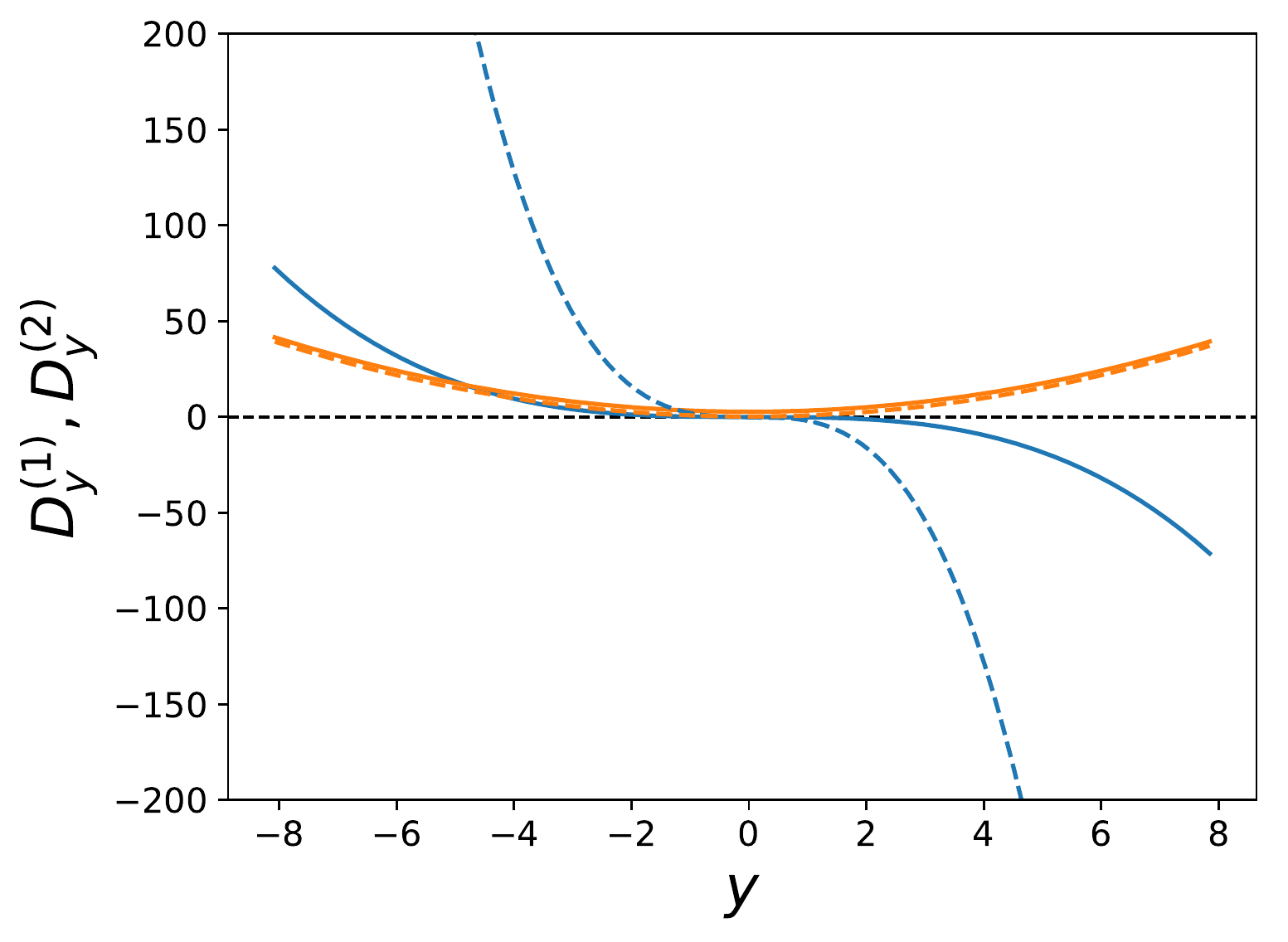}
  \caption{Values of the estimated drift (blue) and diffusion (orange) functions of the first (left) and second (right) component of the HLE, respectively, concerning the synthetic data set generated by the model defined in equation (\ref{HLEsynthetic}). The original functions are plotted as dashed lines, the estimated functions as solid lines. The estimates of $D^{(1)}_x$ and $D^{(2)}_y$ fit the original functions well. The estimates of $D^{(2)}_x$ and $D^{(1)}_y$ deviate from the original ones.\label{fig:D12xysynth}}
\end{figure}

\begin{figure}
  \includegraphics[width=0.32\hsize]{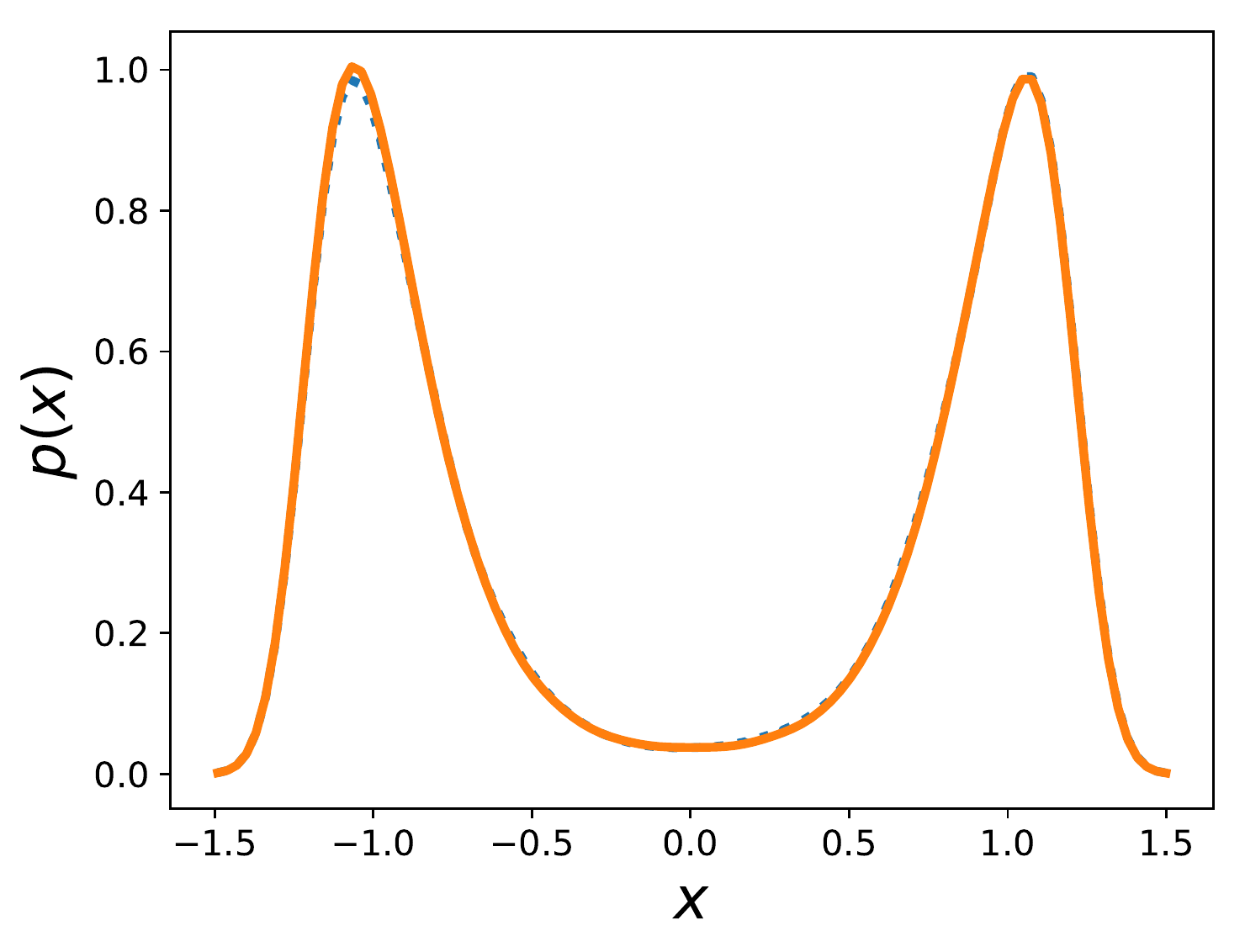}
  \includegraphics[width=0.33\hsize]{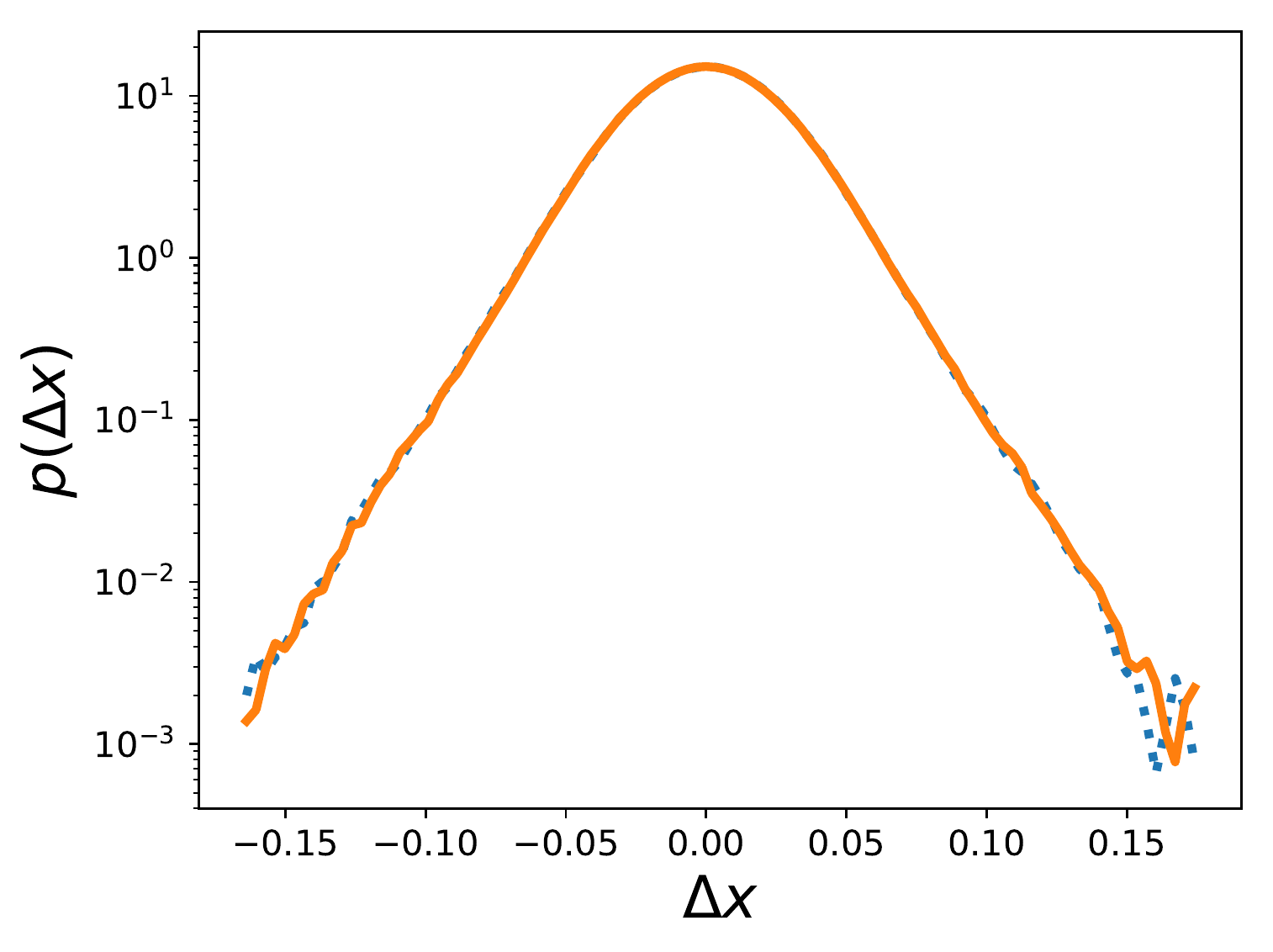}
  \includegraphics[width=0.32\hsize]{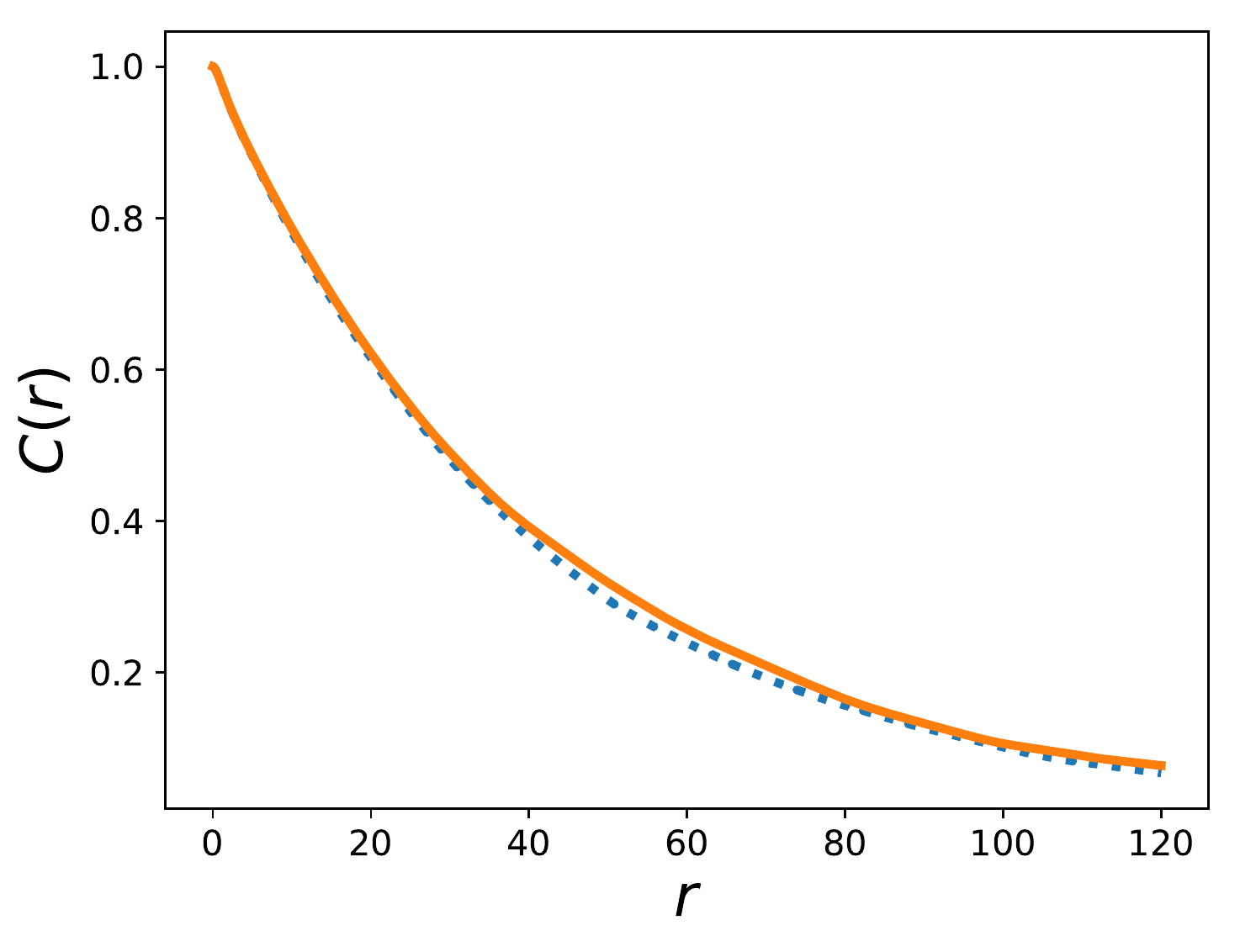}
  \caption{Gaussian kernel-density estimates of the one-point distributions (left), increment distributions (middle, logarithmic plot, $\Delta X_t := X_{t+\Delta t} - X_t$), and autocorrelation functions $C$ (right) of the first component $X_t$ of the estimated (orange, solid) and original (blue, dashed) HLE model, respectively. See equation (\ref{HLEsynthetic}) for the definition of the original model.\label{fig:statQuantsynth}}
\end{figure}

As shown in figure \ref{fig:D12xysynth}, the finally estimated functions $D^{(1)}_x$ and $D^{(2)}_y$ correspond to the original ones, whereas the functions $D^{(2)}_x$ and $D^{(1)}_y$ show significant deviations. Yet, important statistical properties of the process $X_t$ defined by the estimated model -- as the one-point distribution, the increment distribution, and the autocorrelation function (cf. figure \ref{fig:statQuantsynth}) -- indicate that the estimated model is, with respect to the process $X_t$, similar to the original one. Apparently, the result of the estimation is not unique if only the values of the first component are known. The underestimated slope of the function $D^{(1)}_y$ (leading to a wider distribution of the noise) balances the underestimated values of the function $D^{(2)}_x$. Our MAP estimate seems to be one of different (local) optima of the posterior.

\begin{figure}
  \begin{center}
  \includegraphics[width=0.8\hsize]{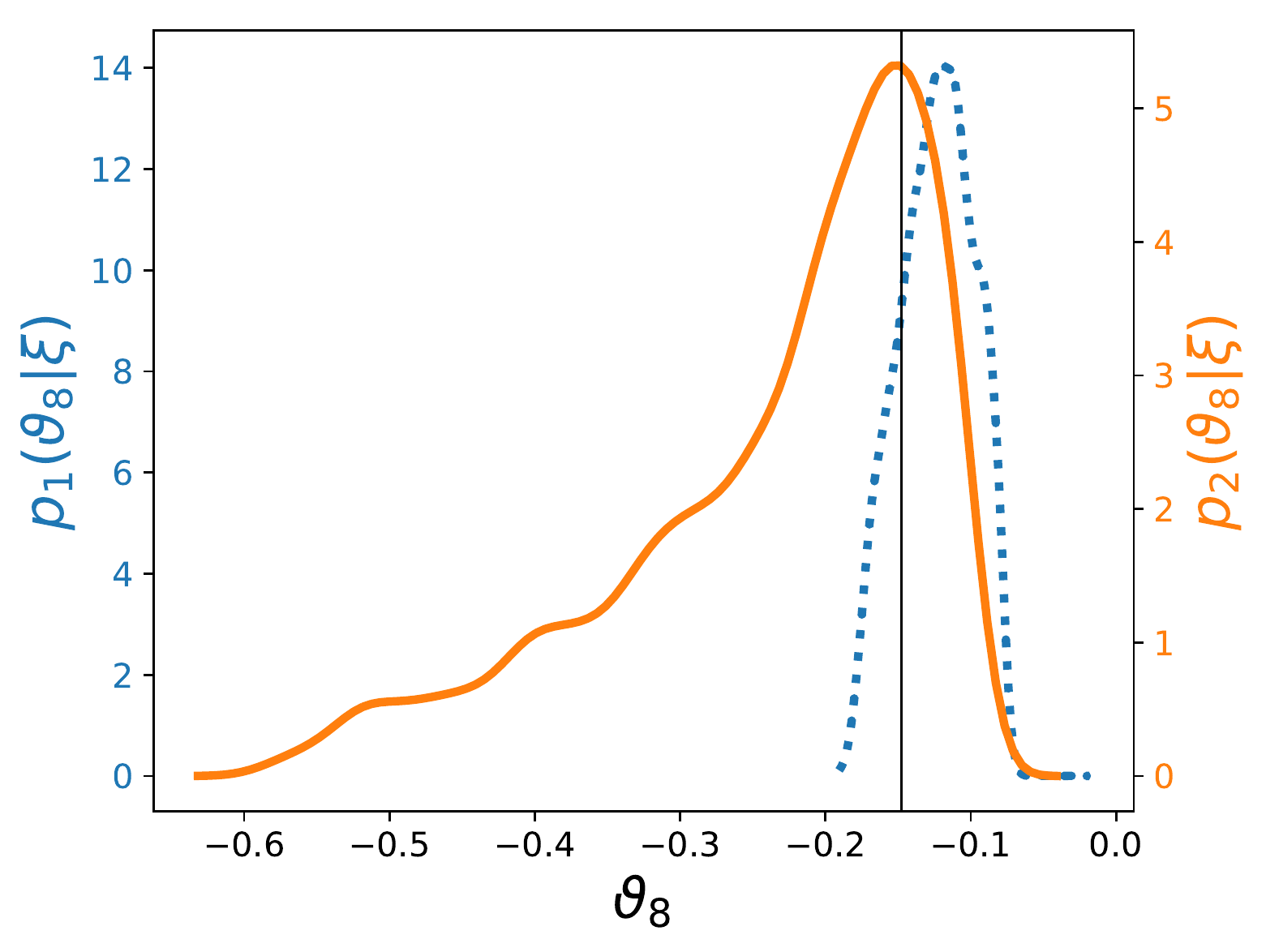}
  \end{center}
  \caption{Marginal posterior distribution of parameter $\vartheta_8$ of the HLE parameterized by polynomials (cf. equation (\ref{ParamSynth})) concerning the synthetic data set generated by the model defined in equation (\ref{HLEsynthetic}). Both versions are calculated by MCMC sampling and Gaussian kernel-density estimation. $p_1$ (blue, dotted) includes $5\cdot 10^5$ samples (without thinning), which took 39 hours on a standard desktop PC. $p_2$ (orange, solid) includes $2.5\cdot 10^6$ samples (without thinning), which took 66 hours with parallel execution of the sampler on the same computer (12 cpu kernels). At reasonable expense, a reliable computation of the posterior distribution is not feasible. The estimated value of $\vartheta_8$ (MAP) is marked by a vertical line.\label{fig:MCMCsynth}}
\end{figure}

Further, the numerical calculation of the distribution of the posterior by MCMC sampling proves problematic in this example (we perform MCMC sampling via python package emcee \cite{a:Foreman-Mackey2013}, which provides an implementation of an affine-invariant ensemble sampler). In figure \ref{fig:MCMCsynth}, we show the exemplary plot of the resulting marginal posterior distribution of the parameter $\vartheta_8$. By generating $5\cdot 10^5$ samples (without thinning; mean autocorrelation time: 590 steps), which took 39 hours on a standard desktop PC, we obtain a single peak whose maximum is located at approximately $-0.14$. Yet, the original value is $-2$. Investing an even greater computational effort ($2.5\cdot 10^6$ samples without thinning; mean autocorrelation time: 2700 steps; 66 hours computation time, parallel on 12 cpu kernels via python module multiprocessing), we see a wider peak. However, at reasonable expense, it is not possible to sample the posterior in such a way that the MCMC algorithm converges. In particular, the mean autocorrelation times of the samples do not converge as well. Consequently, it is unfeasible to obtain reliable confidence intervals. Besides, the ambiguity of the estimation problem cannot be reflected.

The difficulty arising here is that the estimation of the HLE model forms a numerically expensive ill-posed inverse problem. We discuss this topic in more detail in the next section.

\section{Ill-posed expensive vs. well-posed efficient estimation problems}\label{sec:HLE-OU}

Via the posterior evolved in section \ref{sec:Propagator}, Bayesian MAP estimation of the parameters $\vartheta$ of the HLE model is feasible -- even in non-trivial situations as the bimodally distributed data set discussed in section \ref{sec:exampleSynth}. Yet, as seen in this example, the posterior exhibits different optima, which means that the MAP estimate is not unique. The estimation of the HLE model given an observed time series of only the first component forms an ill-posed inverse problem\footnote{In reference \cite{JensenPHD}, this issue is discussed for a similar model in which all components incorporate white noise.}, i.e., there are different models resulting in the same statistical properties of the measured component $X_t$. An analytic argument for this issue is the following example. The noise process of the model can be modified by a linear transformation $\tilde{Y}_t := a + b Y_t$ without changing the observed process $X_t$, if the drift and diffusion functions are adjusted as follows \cite{BerndLinTrafo}:
\begin{subequations}
\label{noiseTrafo}
\begin{align}
\tilde{D}^{(1)}_x(X_t) &:= D^{(1)}_x(X_t) - \frac{a}{b}\sqrt{D^{(2)}_x(X_t)} & \tilde{D}^{(2)}_x(X_t) &:= \frac{1}{b^2}D^{(2)}_x(X_t) \noindent\\
\tilde{D}^{(1)}_y(\tilde{Y}_t) &:= b D^{(1)}_y\left( \frac{\tilde{Y}_t-a}{b} \right) & \tilde{D}^{(2)}_y(\tilde{Y}_t) &:= b^2 D^{(2)}_y\left( \frac{\tilde{Y}_t-a}{b} \right)
\end{align}
\end{subequations}

\noindent Actually, the ambiguity observed in the example in section \ref{sec:exampleSynth} corresponds to the transformation $\tilde{Y}_t = 3.7\, Y_t$, which proves the equivalence of the synthetic model and the estimation result.

The fact, that the estimation problem is ill-posed does not necessarily affect the feasibility of the MAP estimation. It is a difficult (or probably impossible) task to find the global optimum of the posterior, i.e., the best MAP estimate. Yet, also local optima can provide reliable MAP estimates, as we have seen in the example in section \ref{sec:exampleSynth}.

However, the same example indicated that MCMC sampling of the posterior can prove problematic in the sense that it does not converge or reflect the ambiguity of the estimation at reasonable expense. In the case that a single peak can be expected, MCMC sampling facilitates the calculation of a probability distribution even in high-dimensional parameter spaces through exploring the vicinity of the peak only. Yet, it is extremely problematic to sample a high-dimensional parameter space extensively (this is related to the problem of high-dimensional global optimization). Additionally, the above arbitrary scaling of the noise process via the parameter $b$ (cf. equation (\ref{noiseTrafo})) causes that most polynomial parameters of the model (if regarded separately) can take arbitrary values independently of the data. Consequently, the corresponding marginal posterior distributions of individual parameters cannot have the shape of localized peaks (this could be an explanation for the convergence problems in our example) and it is questionable how credible intervals for single parameters could even make sense. Thus, it is problematic to take advantage of all capabilities of Bayesian estimation in the context of this ill-posed estimation problem.

A solution of this issue could be a regularization of the inverse problem of estimating the general HLE model \cite{Tikhonov,jaynes1984prior}. Hereby, physically reasonable assumptions could be involved such as that the time scale of the noise process should not be larger than the time scale of the measured process, which would simplify the interpretation of the model. Yet, further research into this direction is not subject of this work.

Instead, we discuss Bayesian estimation of a specific version of the HLE model. We restrict its degrees of freedom by determining that the noise process $Y_t$ is an Ornstein-Uhlenbeck (OU) process. We refer to this model by the abbreviation HLE-OU (cf. equation (\ref{HLE})). Thereby, we define the noise process by the single parameter $\theta$. This parameter defines the correlation length of the process, whereby the variance is fixed to $\frac12$ \cite{directEst}. This ensures that the affin-linear transformation of the noise mentioned above (cf. equation (\ref{noiseTrafo})) is unfeasible, which considerably reduces the ambiguity of the estimation. According to our experience, the posterior of the HLE-OU is much easier to handle and does not show the problems seen in the general model (see also our previous work \cite{directEst}). Though, a strict proof of well-posedness of the HLE-OU or a similar restricted model is not part of this work.

In addition and even more importantly, we can solve the difficulty of high numerical expense in the case of the HLE-OU model. During both optimization and sampling, a lot of function calls of the posterior have to be carried out. Every calculation of a posterior value involves a sum that is as long as the data set. When dealing with data sets which have a length of $10^5$, $10^6$, or $10^7$ (as we do in our examples), the estimation is numerically very expensive. For MCMC sampling, this can result in a computation time of more than several days on a standard desktop PC (cf. section \ref{sec:exampleSynth}). Also for MAP estimation, the computation takes many hours, if global optimization techniques are employed. In this way, a broad test of different parameterizations of the drift and diffusion functions is unfeasible. 

Now, for the HLE-OU model, it is possible to perform the calculation of the posterior in a numerically efficient way. Hereby, dealing with large data sets is no longer a problem. We describe this approach in the next two sections.

\section{Direct estimate of SLE as initial guess for HLE-OU}\label{sec:directEstSLE}

To obtain an initial guess for the estimation of the HLE-OU (cf. equation (\ref{HLE})), we combine the drift and diffusion functions of an estimated SLE with a small value for the parameter $\theta$. This parameter defines the correlation length of the HLE-OU noise process \cite{directEst}. Thus, a small value (compared to the correlation length of the observed process) leads to noise which is similar to the uncorrelated noise of the SLE. The drift function of the SLE provides an initial guess for the function $D^{(1)}_x$ of the HLE-OU. The diffusion function has to be multiplied by the factor $\sqrt{2}$, because the variance of the noise $Y_t$ of the HLE-OU equals $2$ (independendly of the parameter $\theta$) \cite{directEst}, whereas the variance of the uncorrelated noise $\eta_t$ of the SLE is $1$.

The most important estimation methods for the SLE are nonparametric \textit{direct estimation}, which is very efficient and was introduced by Friedrich et al. \cite{directEstFriedrich,directEstSiegert,DEcorrection0,DEcorrection1,DEcorrection2,ReviewFriedrichPeinke}, and parametric maximum likelihood estimation \cite{Kleinhans,KLEINHANS_MLE}. The latter can easily be extended to parametric Bayesian estimation (cf. section \ref{sec:Bayes}). Nonparametric Bayesian estimation methods are discussed in reference \cite{NonParBayes_SLE}. Direct estimation of the SLE is strongly connected with the efficient way of Bayesian estimation of the HLE-OU. Therefore, we choose this method to obtain an effective initial guess. We briefly introduce this technique in the following.

Direct estimation is often called a \textit{nonparametric} method. However, our understanding is that direct estimation is the estimation of the drift and diffusion functions parameterized as piecewise constant functions. Precisely, for a measured time series $\xi=(x_0, x_1, ..., x_{N_{\text{data}}})$ that is sampled with a time step $\Delta t$, the range of the measured values $D = \lbrack x_{\min}, x_{\max} \rbrack$ is divided into $N_{\text{bins}}$ disjoint parts $B_j$: $D = \cup_{j=1}^{N_{\text{bins}}} B_j$. For every $B_j$, an averaged function value for the drift and diffusion function is estimated as follows:

\begin{subequations}
\label{directEst}
\begin{align}
\hat{D}^{(1)}_j &= \lim_{\Delta t\to 0} \frac{1}{\Delta t} \langle x_{i+1} - x_i | x_i\in B_j \rangle\\
\hat{D}^{(2)}_j &= \lim_{\Delta t\to 0} \frac{1}{\Delta t} \Bigl\{ \langle (x_{i+1} - x_i)^2 | x_i\in B_j \rangle \nonumber\\
&\qquad\qquad\qquad - \lbrack \hat{D}^{(1)}_j\rbrack^2\,\Delta t^2 \Bigr\}
\end{align}
\end{subequations}

\noindent The limit $\Delta t\to 0$ can be achieved by a polynomial extrapolation of the values obtained for the time steps $\tau$, $2\tau$, $3\tau$, etc. Here, we use the single value for the time step $\Delta t$ as an approximation of 0th order.

\section{Efficient Bayesian estimation of HLE-OU model}\label{sec:Efficient}

In the case of the HLE-OU model (cf. equation (\ref{HLE})), the numerical calculation of the posterior -- and thereby the whole estimation process -- can be implemented much more efficiently, if the drift and diffusion functions are parameterized as piecewise constant functions. Hereby, the efficiency of direct estimation is combined with the capabilities of Bayesian estimation. This idea was previously used by Kleinhans \cite{Kleinhans} for the SLE. In the following, we apply it to the HLE-OU model.

First, for the values of a measured time series $\xi=(x_0, ..., x_{N_{\text{data}}})$ with the sampling time step $\Delta t$, we perform a binning $\lbrack x_{\min}, x_{\max} \rbrack = \cup_{j=1}^{N_{\text{bins}}} B_j$ as described in the context of direct estimation in Sec.~\ref{sec:directEstSLE}.  The number of bins $N_{\text{bins}}$ should be neither too small nor too great. If it is too small, the resulting functions are not smooth enough. If it is too great, statistical errors come into account, because every bin contains too few data. The distribution of the bins can be even or uneven depending on the properties of the data set. An uneven binning can be useful, for example, if the distribution of the measured values shows a sharp edge on only one side. In this case, we would expect that the curves of the drift and diffusion functions require a very detailed resolution on the same side. Second, we define the piecewise constant parameterization of the drift and diffusion functions $D^{(1)}_x$ and $D^{(2)}_x$ correspondingly:

\begin{align}
D^{(1)}_x(x) &:= D^{(1)}_j, \quad x\in B_j\\
D^{(2)}_x(x) &:= D^{(2)}_j, \quad x\in B_j.
\end{align}

\noindent Now, the parameters of the HLE-OU are $\vartheta=(D^{(1)}_1, ..., D^{(1)}_{N_{\text{bins}}}, D^{(2)}_1, ..., D^{(2)}_{N_{\text{bins}}},\theta)$.

Next, we rearrange the terms of the sum which occurs in the logarithm of the likelihood (cf. equation (\ref{likelihood_unsorted})) accordingly through replacing the sum $\sum_{i=1}^{N_{\text{data}}-1}$ by the double sum $\sum_{j=1}^{N_{\text{bins}}} \sum_{x_i,x_{i-1}\in B_j}$. Afterwards, by employing the parameterization of the functions $D^{(1)}_x$ and $D^{(2)}_x$, we replace the terms $D^{(1)}_x(x_i)$, $D^{(1)}_x(x_{i-1})$, $D^{(2)}_x(x_i)$, and $D^{(2)}_x(x_{i-1})$ by $D^{(1)}_j$, and $D^{(2)}_j$, respectively.

With the abbreviation

\begin{equation}
\Delta_i := x_i - x_{i-1}
\end{equation}

\noindent and several reformulations, we obtain

\begin{multline}
\log p(\xi|\vartheta) = \sum_{j=1}^{N_{\text{bins}}} \sum_{x_i,x_{i-1}\in B_j} \Biggl\{ \alpha_j^0  + \alpha_j^1\Delta_{i+1} \\
\qquad + \alpha_j^2\Delta_i + \alpha_j^3\Delta_{i+1}^2 + \alpha_j^4\Delta_{i+1}\Delta_i + \alpha_j^5\Delta_i^2 \Biggr\},
\end{multline}

\noindent where the model parameters are to be found in the following terms:

\begin{align}
\alpha_j^0 &:= -\frac12\log\left(2\pi D_j^{(2)}\frac{1}{\theta}\Delta t^3\right) - \frac{\left(D_j^{(1)}\right)^2\Delta t}{2D_j^{(2)}\theta} \nonumber\\
\alpha_j^1 &:= \frac{D^{(1)}_j}{D^{(2)}_j\Delta t} \nonumber\\
\alpha_j^2 &:= \frac{D^{(1)}_j}{D^{(2)}_j}(\frac{1}{\theta} - \frac{1}{\Delta t}) \nonumber\\
\alpha_j^3 &:= -\frac{\theta}{2D^{(2)}_j\Delta t^3} \nonumber\\
\alpha_j^4 &:= \frac{\theta - \Delta t}{D^{(2)}_j\Delta t^3} \nonumber\\
\alpha_j^5 &:= -\frac{(\Delta t - \theta)^2}{2D^{(2)}_j\theta \Delta t^3}.
\end{align}

Finally, with the definition

\begin{equation}\label{coefficients}
c^{nm}_j := \sum_{x_i,x_{i-1}\in B_j} \left(\Delta_{i+1}\right)^n \left(\Delta_i\right)^m
\end{equation}

\noindent we end up with a single summation:

\begin{align}\label{likelihood_sorted}
\log p(\xi|\vartheta) = \sum_{j=1}^{N_{\text{bins}}} \Biggl\{ \alpha_j^0c_j^{00}  + \alpha_j^1c^{10}_j  + \alpha_j^2c^{01}_j + \alpha_j^3c^{20}_j + \alpha_j^4c^{11}_j + \alpha_j^5c^{02}_j \Biggr\}
\end{align}

Here, the coefficients $c^{nm}_j$ ($j\in\{1,...,N_{\text{bins}}\}$, $n,m\in\{0,1,2\}$, $n+m \le 2$) represent all information of the data set $\xi$ that is relevant to Bayesian estimation of the HLE-OU model. At the same time, they include the majority of floating-point operations (FLOPS) needed for an evaluation of the likelihood. Thus, the advantage of the reformulation of the likelihood function is that the coefficients $c^{nm}_j$ can be calculated before the optimization is carried out. This reduces the number of FLOPS needed for the estimation, which includes repeated evaluations of the likelihood function, considerably. In this way, the estimation is efficient also for large data sets. The numerical cost of optimizing and sampling the posterior even gets independent of the length of the data set. To give an idea of the extent of this effect, we list the lengths of the different sums occuring in the equations (\ref{likelihood_unsorted}), (\ref{coefficients}), and (\ref{likelihood_sorted}) in the case of the free jet turbulence example which we discuss in section \ref{sec:exampleTurb}: $10^7$ ($\sum_{i=1}^{N_{\text{data}}-1}$), $10^6$ ($\sum_{x_i,x_{i-1}\in B_j}$), and $10$ ($\sum_{j=1}^{N_{\text{bins}}}$). So, the length of the sum occuring in the likelihood is reduced from $10^7$ to $10$.

This advantage in computational cost is not restricted to the piecewise constant parameterization of the functions $D^{(1)}_x$ and $D^{(2)}_x$. It can be used for arbitrary parameterizations, if these are approximated in the piecewise constant way in each calculation of the posterior \cite{KLEINHANS_MLE}. Further, if necessary, the prior could be modified such that it favors smooth curves of the functions \cite{Hummer_2005}.

\section{Example of HLE-OU model estimation: free jet turbulence\label{sec:exampleTurb}}

Turbulent flows appear in many different examples in nature and technology. Because of their spatio-temporal complexity, their modeling is still a challenge. A fundamental concept in turbulence research is the cascade model \cite{Richardson,Frisch}. It describes qualitatively how spatial turbulent structures evolve: They emerge on different length scales and are instable according to the nonlinearity of the Navier-Stokes equation. Large structures decompose into smaller structures whereby the corresponding kinetic energy is transferred. This process of structure decomposition and energy transfer proceeds stepwise from the largest structures, that are initiated by external forces, to the smallest structures, that are determined by viscosity and in which energy is dissipated. There is a specific range of scales in which it can be assumed that the cascade process is self-similar and neither affected by external forces nor by dissipation. This range is called the inertial range and bounded by two length scales. The upper bound is the correlation length $L$. The lower bound is the Taylor length $\Lambda$. Both length scales are based on the curve of the autocorrelation function (ACF) $C(r) = (\langle u(r_0+r)u(r_0) \rangle - \langle u(r_0)\rangle^2)/\langle u(r_0)^2 \rangle$ of the longitudinal velocity $u(r)$. A spatial velocity series $u(r)$ can be obtained by a measurement of a temporal velocity series $u(t)$ at a fixed location and a transformation according to the Taylor hypothesis \cite{Turb:Reinke}.

An important experimental setup in which velocity series are measured is a free jet \cite{Turb:Chanal,Turb:Reinke}. Hereby, air (or helium, etc.) streams out of a nozzle with diameter $d$ and the longitudinal velocity is measured at a downstream position $x$ by, e.g., hot-wires. Via the pressure of the gas, the Reynolds number of the free jet can be adjusted in a wide range.

Here, we employ the HLE-OU and the efficient estimation technique described in this work to model the longitudinal velocity of a free jet turbulence. Thereby, our goal is to reproduce the original ACF, which is an important statistical quantity of the turbulent data as explained above.

The data that we use comes from group (v) in reference \cite{Turb:Reinke}. It is a free jet in air with a ratio of $x/d=40$ and a Taylor-Reynolds number of $865$. The time series contains $1.6\cdot 10^7$ data points. The distance of two samples corresponds to a length of $1.31\cdot 10^{-4}\,\text{m}$. The unit of the measured velocities is $\frac{\text{m}}{\text{s}}$. For the sake of simplicity, we omit all units in the following. We calculate a correlation length that corresponds to roughly 1480 samples and a Taylor length that corresponds to roughly 60 samples. We investigate this time series by taking every 200th sample into account which corresponds to a length scale that is located in the lower region of the inertial range. This implies that the data is non-Markovian.

For the sake of simplicity, we choose $\Delta t = 1$. Changing this value would only result in a scaling of the model parameters (effectively, we treat the HLE-OU as a time-discrete model via its Euler-Maruyama approximation).

To estimate the HLE-OU model, we first use direct estimation to find an appropriate SLE model. Thereby, we determine a suitable distribution of the bins $B_j$, which leads to drift and diffusion functions that are both sufficiently smooth and not affected by large statistical errors. We choose $N_{\text{bins}}=10$ evenly distributed bins.

Further, in this example, it is necessary to neglect the highest and lowest values of the data set. These would lead to very sparsely filled bins in the outer regions of the domain of the drift and diffusion functions. The statistical errors induced by them would cause problems during optimization of the posterior of the HLE-OU model. We spread the 10 bins within the interval $\lbrack -5.5, 5.5\rbrack$ and neglect all data points outside this interval ($u_{\text{min}}=-7.27$, $u_{\text{max}}=10.31$). Hereby, we only loose 0.6\% of the data. Later, we extrapolate the drift and diffusion functions in a suitable way.

Next, we use the SLE estimate to generate an initial guess for the HLE-OU model as described in section \ref{sec:directEstSLE} (with an initial value of $0.001$ for $\theta$). On the basis of this initial guess, we optimize the posterior of the HLE-OU model as developed in section \ref{sec:Efficient} to obtain an estimate for the model in the sense of the piecewise constant parameterization of the functions $D^{(1)}_x$ and $D^{(2)}_x$ and the parameter $\theta$. For the numerical optimization, we apply a combination of the Powell \cite{Powell} and Nelder-Mead \cite{NelderMead} algorithms as in the first example in this work (section \ref{sec:exampleSynth}).

Finally, we perform an MCMC analysis to obtain marginal posterior distributions of the individual parameters. Again, we use the python package emcee \cite{a:Foreman-Mackey2013}, which provides an implementation of an affine-invariant ensemble sampler, and generate starting parameters through the MAP estimates. The sampling took 16 minutes for $2\cdot 10^7$ samples (without thinning) on a standard desktop PC. The autocorrelation time of the sample chain is roughly $400$. Thus, we obtain an effective number of $50000$ samples. Comparing this computational cost to the cost that occurred in the example discussed in section \ref{sec:exampleSynth}, we see the significant advantage of the efficient Bayesian estimation developed in section \ref{sec:Efficient}.

\begin{figure}
  \begin{center}
  \includegraphics[width=0.9\hsize]{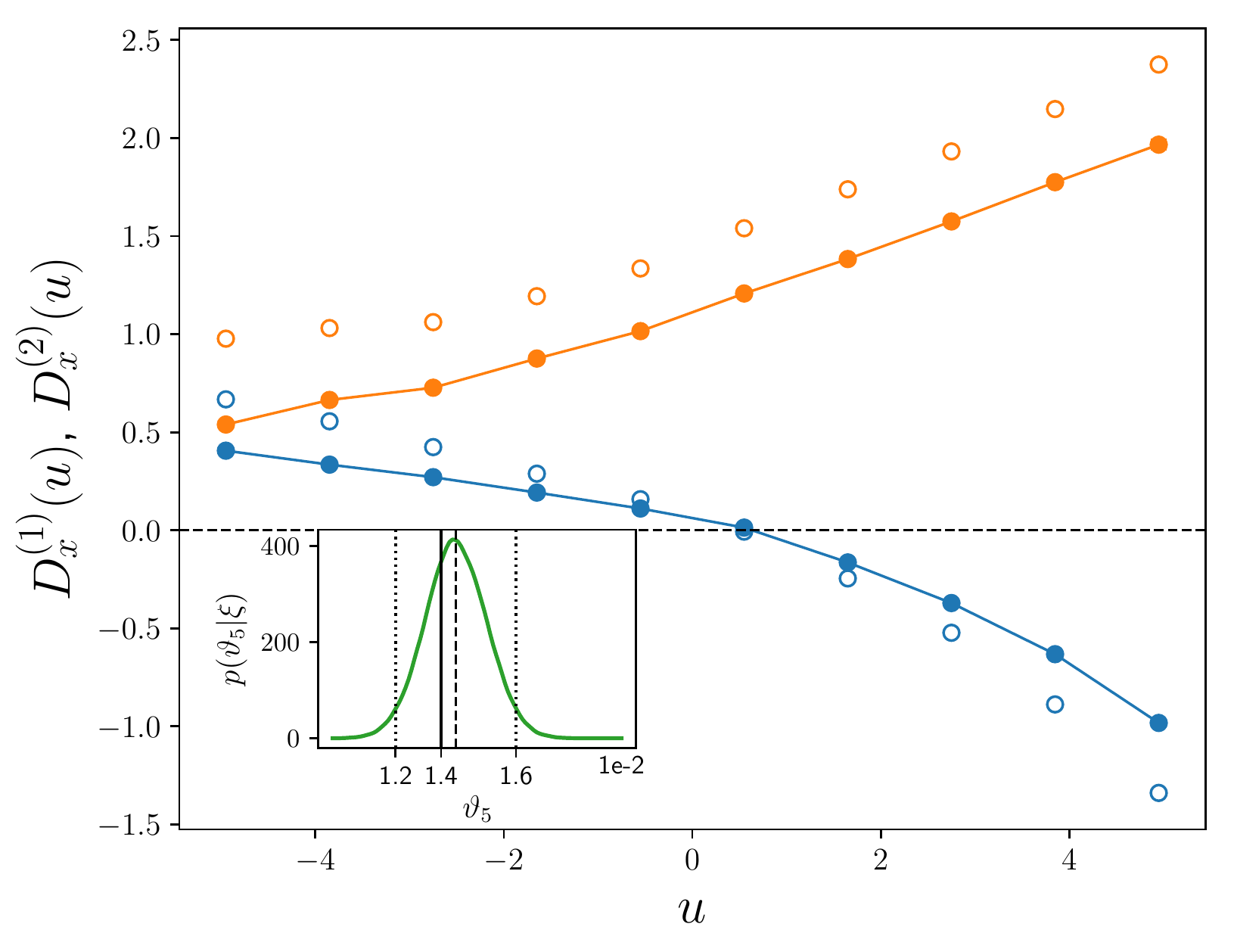}
  \end{center}
  \caption{Values of the estimated drift (blue) and diffusion (orange) functions of the SLE (open circles; used to create an initial guess for the HLE-OU) and the HLE-OU at the centers of the bins with linear interpolations (filled circles and lines) of the free jet turbulence. Due to the length of the data set, all credible intervals are too small to be visible as error bars in this plot. As an example, the inset shows the marginal posterior distribution of the parameter $\vartheta_5$ (close to the zero of $D^{(1)}_x$) with estimated value, 95\% credible interval, and mean marked by vertical lines. \label{fig:D12}}
\end{figure}

In figure \ref{fig:D12}, we show the estimated drift and diffusion functions of the HLE-OU model. The estimated SLE provides an effective initial guess which is close to the final estimation result. Due to the length of the data set, credible intervals are too small to be visible as error bars. The estimated value of the correlation length of the noise process is $\theta=0.84$. The exemplary plot of a marginal posterior distribution indicates that the MAP estimation result deviates slightly from the mean of the posterior. The latter provides a marginally different estimate, which, due to the sharpness of the posterior, leads to the same curve of the ACF (not shown).

For the integration of the estimated model (cf. equation (\ref{EulerMaruyama})), we interpolate the function values of $D^{(1)}_x$ and $D^{(2)}_x$ linearly between the bins while the function values at the centers of the bins are determined by the estimated parameters. Here, also higher order interpolations would be possible. Outside the interval $\lbrack -5.5, 5.5\rbrack$, we extrapolate the functions constantly according to the leftmost and rightmost estimated parameters. This is the simplest extrapolation which definitely does not add any change of sign to the estimated drift and diffusion functions. This is important, because an additional change of sign in the drift function induces another (instable) fixed point, which would significantly affect the dynamics of the estimated model. A change of sign in the diffusion functions would disagree with the assumption that these are positive functions.

\begin{figure}
  \begin{center}
  \includegraphics[width=0.9\hsize]{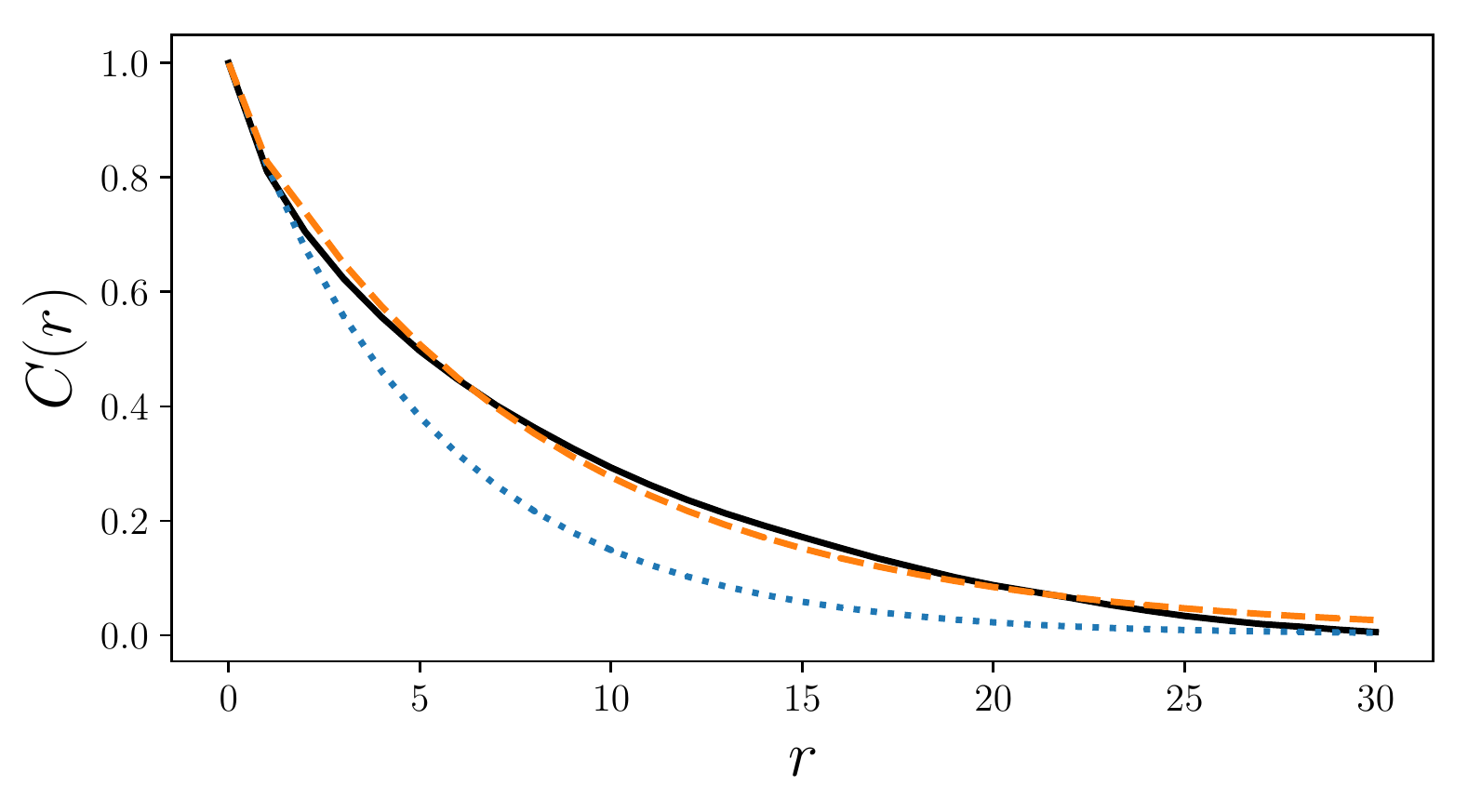}
  \end{center}
  \caption{Autocorrelation function of original and simulated time series of the free jet turbulence. Original data (black solid), SLE model (blue dotted), and HLE model (orange dashed). The unit of the $r$-axis is the spacing between two consecutive samples. \label{fig:ACF}}
\end{figure}

In figure \ref{fig:ACF}, we plot the ACF of the estimated model (via a time series obtained by integration). The plot shows a good accordance with the ACF of the original time series, which we aimed to reproduce. At the same time, by means of the ACF, the estimated, non-Markovian HLE-OU model significantly outperforms the estimated Markovian SLE model.

\begin{figure}
  \includegraphics[width=0.5\hsize]{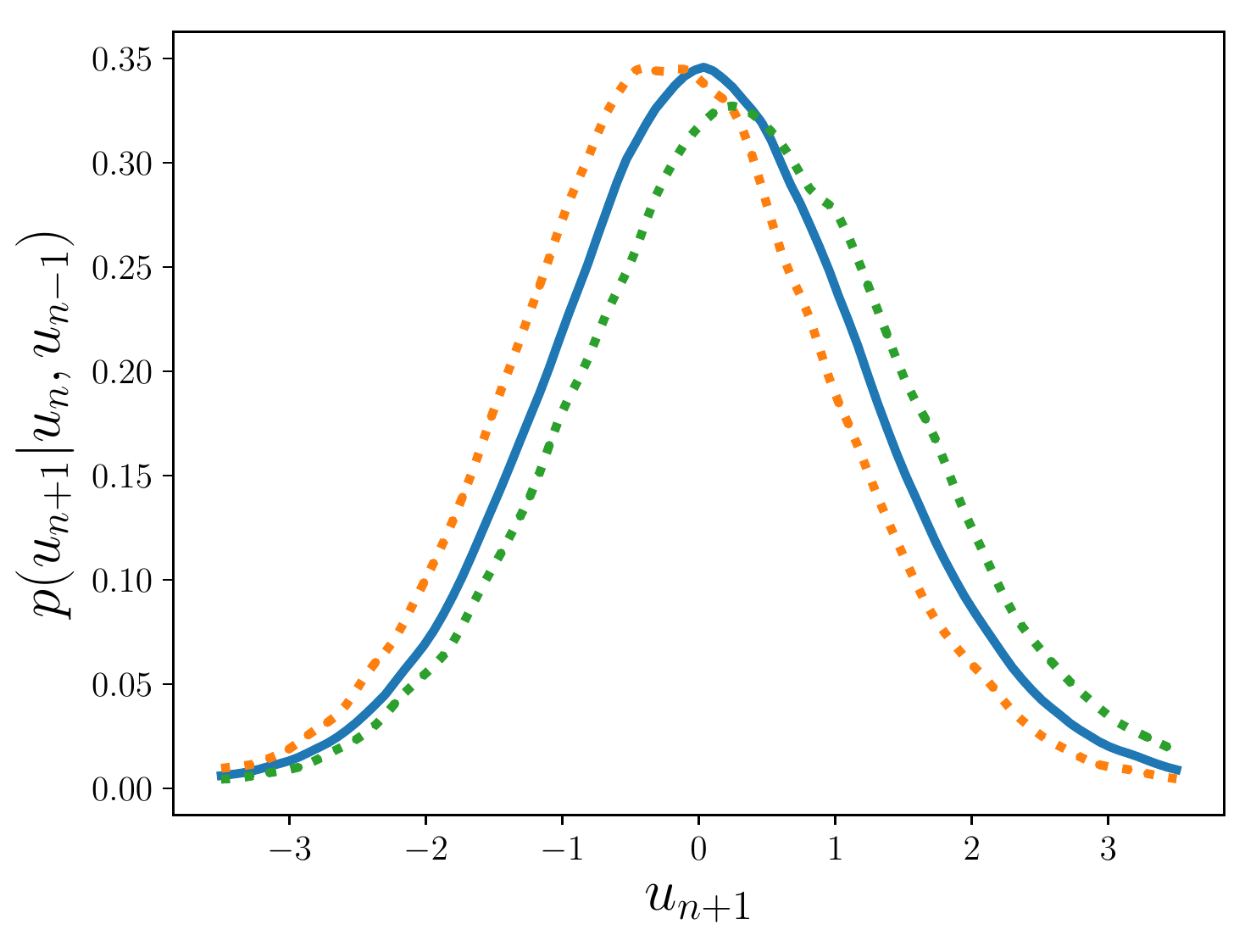}\includegraphics[width=0.5\hsize]{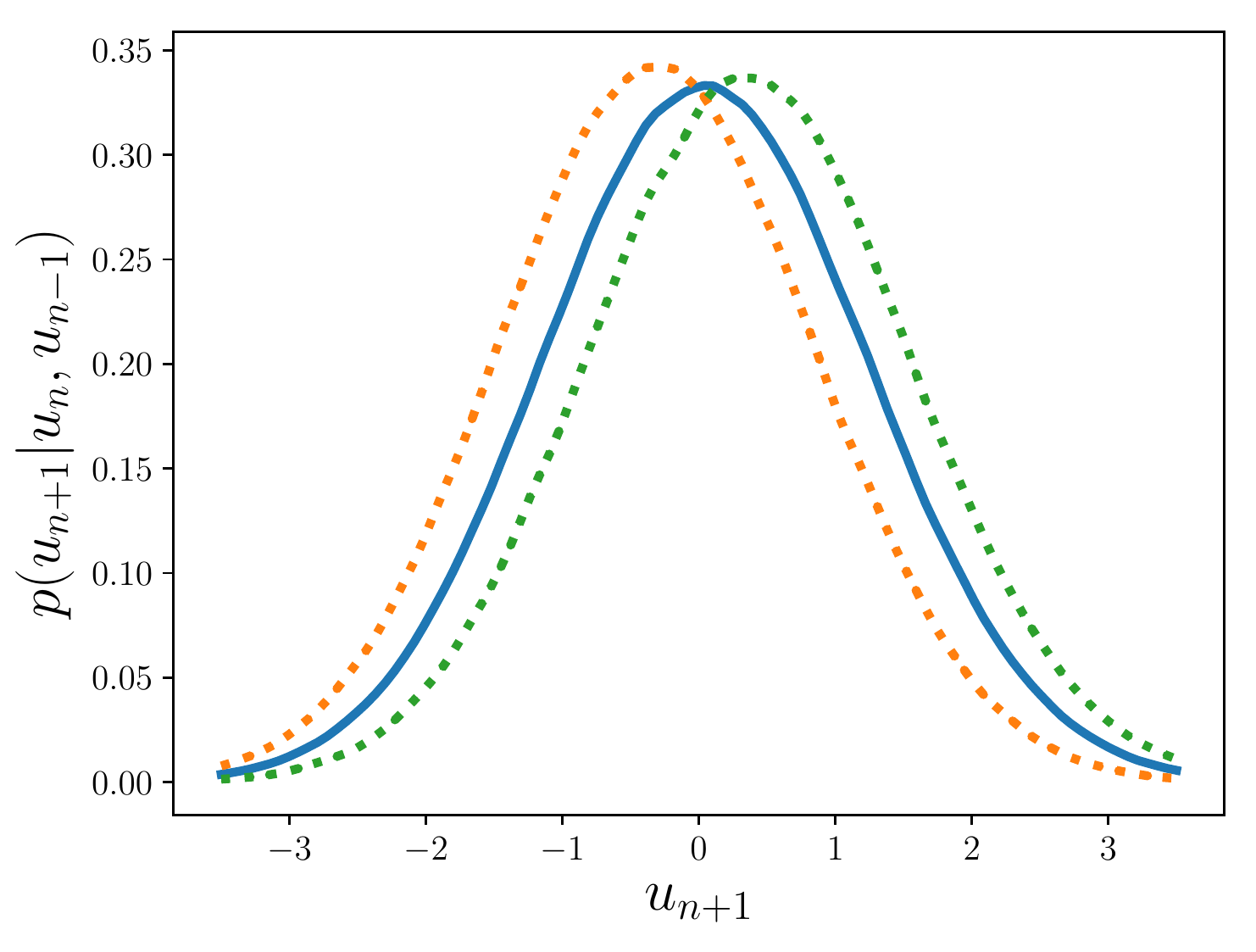}
  \caption{Conditional probability distributions $p(u_{n+1}|u_n\in I_1)$ (solid, $I_1=\lbrack -0.5, 0.5\rbrack$) and $p(u_{n+1}|u_n\in I_1, u_{n-1}\in I_0)$ (dashed, $I_0=\lbrack -1,3\rbrack$ or $I_0=\lbrack -3,1\rbrack$) of the original free jet turbulence data (left) and the estimated HLE-OU model (right).\label{fig:nonMarkovianity}}
\end{figure}

To further illustrate the non-Markovianity of the data and the estimated HLE-OU model, we plot distributions $p(u_{n+1}|u_n\in I_1, u_{n-1}\in I_0)$ for different choices of $I_0$ in figure \ref{fig:nonMarkovianity}. According to the non-Markovianity, the plot shows differences between these distributions. Thereby, the estimated HLE-OU model reproduces the original distributions up to a few details.

All in all, modeling the present data set requires a non-Markovian model. Here, the HLE-OU model, as estimated by the method described in this work, is able to reproduce important statistical properties of the free jet turbulence.

\section{Conclusion and outlook}\label{sec:summary}

In this work, we have derived the posterior distribution of a non-Markovian Langevin model with a hidden, correlated noise process (the HLE model, cf. equation (\ref{HLEgeneral})) via its short-term propagator, which we have evaluated by marginalization. In the sense of Bayesian statistics, this allows the calculation of an MAP estimate of the HLE model, given a measured time series, which we have demonstrated in a synthetic example.

Via both the example and an analytical argument, we have shown that the result of an estimation of the HLE model is not unique, i.e., the estimation is an ill-posed inverse problem. Hence, MAP estimation yields a local maximum of the posterior, which is not necessarily the best one. Here, so-called global optimization techniques -- such as basin-hopping, simulated annealing, or differential evolution -- could be considered to improve the MAP estimate. Also, a different initial guess -- as that one used for the HLE-OU in section \ref{sec:exampleTurb} -- could lead to an alternative estimate. Yet, finding the best MAP estimate, i.e. the global maximum of the posterior, remains a difficult task, especially because an evaluation of the posterior is numerically expensive in the context of large data sets, which we intend to investigate. For the same reasons, MCMC sampling of the posterior turned out to be difficult. In our example, it was not possible to obtain a reliable MCMC estimate of the posterior distribution at reasonable expense. Further, as we discussed in section \ref{sec:HLE-OU}, the idea of credible intervals for single parameters is problematic in the case of this ill-posed estimation problem.

We have circumvented these issues by a restriction of the degrees of freedom of the HLE model. If the noise is an OU process defined by only one parameter (which leads to the HLE-OU model, cf. equation (\ref{HLE})), the ambiguity of the estimation is considerably reduced. At the same time, through a piecewise constant parameterization of the drift and diffusion functions of the first component of the model, it is possible to calculate the posterior in a much more efficient way. This approach can be used for arbitrary parameterizations as well, if they are approximated in the piecewise constant manner in every calculation of the posterior. In this way, stable results of both MAP and MCMC estimation have turned out to be feasible and affordable for the restricted model even in the context of large data sets. Thereby, according to our experience, alternative (global) optimization techniques are not necessary for MAP estimation of the HLE-OU, because they do not cause significantly different results. We have illustrated the procedure by an example from turbulence, in which the non-Markovian model clearly outperforms the Markovian one in terms of the curve of the ACF.

As mentioned earlier, a treatment of the ill-posedness of the estimation of the general model via regularization could be part of future work. Besides, from a theoretical point of view, the occurence of uncertainties during numerical optimization and sampling of the posterior of the HLE-OU also cannot be excluded, until the well-posedness of this estimation problem is strictly proven. Further, at this point, the general theoretical question arises in what sense the determination of credible intervals is to be understood in the case of an estimation problem whose well-posedness is not verified and whose parameter space cannot be explored completely due to limits of computation. This situation might apply to quite a few applications.

Another interesting  question for further research is whether an efficient calculation of the posterior of the general HLE model could be possible as well. This would facilitate an improvement of MAP estimation via global optimization techniques. Here, a crucial problem of the binning approach used for the HLE-OU model would be that the sorting of the $Y$-values into bins has to be performed repeatedly in every calculation of the posterior, because every new parameter set $\vartheta$ implies new $Y$-values. Also the range of the $Y$-values changes for different parameter sets $\vartheta$, which makes repeated adaptations of the bin distribution necessary. We suppose that, due to these problems, the general model is by far not as efficient to estimate as the specific model with the approach employed in this work.

As discussed in section \ref{sec:intro}, both the HLE and the HLE-OU model are applicable in various fields. In turbulence, it could also be employed to the analysis of the turbulent cascade in scale. The dynamics of velocity increments between different scales can be modeled as Markov processes if the jump length between scales is larger than the Taylor length $\Lambda$ (see references \cite{Friedrich1997prl, Renner2001jfm}). Through the method presented in this work, the description in scale might also be possible for smaller jumps.\\

An exemplary python implementation of the proposed method is openly available \cite{Zenodo_HLE}.

\section*{Acknowledgments}

The authors thank J. Peinke and N. Reinke for providing the data set of the free jet turbulence measurement. Further, we thank B. Lehle for helpful discussions concerning the HLE model.

\providecommand{\newblock}{}

\end{document}